\documentclass[aip,jcp,reprint,superscriptaddress]{revtex4-2}%
\usepackage{graphicx}
\usepackage{tabularx}
\usepackage{dcolumn}
\usepackage{longtable}
\usepackage{tensor}
\usepackage{placeins}
\usepackage{bm}
\usepackage{amsmath}
\usepackage{amsfonts}
\usepackage{amssymb}
\usepackage[english]{babel}
\usepackage{gensymb}
\begin{document}

\title{Can increasing the size and flexibility of a molecule reduce decoherence and prolong charge migration?}

\author{Alan Scheidegger}
\email{alan.scheidegger@epfl.ch}
\affiliation{Laboratory of Theoretical Physical Chemistry, Institut des Sciences et
Ing\'enierie Chimiques, Ecole Polytechnique F\'ed\'erale de Lausanne (EPFL),
CH-1015, Lausanne, Switzerland}

\author{Nikolay V. Golubev}
\email{ngolubev@arizona.edu}
\affiliation{Department of Physics, University of Arizona, 1118 E. Fourth Street, Tucson, AZ 85721}

\author{Ji\v{r}\'{\i} J. L. Van\'{\i}\v{c}ek}
\email{jiri.vanicek@epfl.ch}
\affiliation{Laboratory of Theoretical Physical Chemistry, Institut des Sciences et
Ing\'enierie Chimiques, Ecole Polytechnique F\'ed\'erale de Lausanne (EPFL),
CH-1015, Lausanne, Switzerland}

\date{\today}

\begin{abstract}
Coherent superposition of electronic states, created by ionizing a molecule, can initiate ultrafast dynamics of the electron density. Correlation between nuclear and electron motions, however, typically dissipates the electronic coherence in only a few femtoseconds, especially in larger and more flexible molecules.  We, therefore, use ab initio semiclassical dynamics to study decoherence in a sequence of analogous organic molecules of increasing size and find, surprisingly, that extending the carbon skeleton in propynal analogs slows down decoherence and prolongs charge migration. To elucidate this observation, we decompose the overall decoherence into contributions from individual vibrational modes and show that: (1) The initial decay of electronic coherence is caused by \emph{high- and intermediate-frequency} vibrations via \emph{momentum} separation of nuclear wavepackets evolving on different electronic surfaces. (2) At later times, the coherence disappears completely due to the increasing \emph{position} separation in the \emph{low-frequency} modes. (3) In agreement with another study, we observe that only normal modes that preserve the symmetry of the molecule induce decoherence. All together, we justify the enhanced charge migration by a combination of increased hole-mixing and the disappearance of decoherence contributions from specific vibrational modes---CO stretching in butynal and various H rockings in pentynal.
\end{abstract}

\maketitle

Tailored laser pulses can excite a molecule to a superposition of electronic states and trigger an ultrafast electronic oscillation with a period inversely proportional to the energy gap. In some cases, this electronic oscillation is accompanied by changes in the electron density, which can result in the creation~\cite{Valentini_Remacle:2020} or rupture~\cite{Calegari_Nisoli:2014} of chemical bonds. The idea that chemical reactivity of a molecule can be manipulated by photoexcitation laid the foundation for the new field of attochemistry~\cite{Nisoli:2019,Remacle_Weinkauf:1999}.

A prototypical scenario starts with the ionization of a molecule. Under certain conditions, this process can induce oscillations of the hole along the molecular structure. As demonstrated by Weinkauf \textit{et al.}~\cite{Weinkauf_Schlag:1995}, the time dependence of the charge location can be exploited by using a second laser pulse to break the molecule at different sites, depending on the time delay between the two pulses. This type of charge dynamics is driven purely by electron correlation and has been termed ``charge migration"~\cite{Cederbaum_Zobeley:1999,Remacle_Levine:2006} to distinguish it from charge transfer~\cite{Sun_Remacle:2017,Lehr_Schlag:2005}, which is induced by nuclear motion. Nevertheless, charge migration is also strongly coupled to the motion of the nuclei, since their rearrangement after photoexcitation will typically lead to ultrafast decoherence of the electronic oscillation on the femtosecond time scale~\cite{Vacher_Robb:2015,Arnold_Santra:2017,Despre_Kuleff:2018,Matselyukh_Worner:2022}. Theoretical studies often neglect the effect of the nuclei and focus on charge migration within a fixed molecular geometry~\cite{Folorunso_Lopata:2021,Folorunso_Lopata:2023,Despre_Kuleff:2019,Khalili_Shokri:2021,Lunnemann_Cederbaum:2008}. For instance, it has been shown that hole oscillation in parafunctionalized bromobenzenes can be modulated by the electron-donating or -withdrawing character of the substituents~\cite{Folorunso_Lopata:2023}. Investigations of alkenes and alkynes with varying chain lengths demonstrated similar hole dynamics but identified systematic variations depending on the number of carbon atoms~\cite{Folorunso_Lopata:2021, Despre_Kuleff:2019}. Charge migration can also be enhanced by placing the molecule in an optical cavity~\cite{Gu_Mukamel:2023}. Together, these findings suggest that molecules can be tailored to exhibit specific electron dynamics after photoexcitation, which is an essential step in the development of compounds for attochemistry applications.

If nuclear motion is included, the electronic coherence can be suppressed in just a few femtoseconds or, in contrast, it can last for several tens of femtoseconds~\cite{Scheidegger_Vanicek:2023}.
Simulations showed that increasing the mass of the nuclei may reduce the decoherence rate~\cite{Hu_Franco:2018}, consistent with the experimental observation of longer coherence in DBr compared to HBr~\cite{Kobayashi_Leone:2020}. Isotopic substitution conserves chemical reactivity, but offers limited possibilities. In contrast, increasing the mass of a molecule by altering its atomic composition can be done in many ways, but generally changes its chemical properties. Based on observations from spin chains~\cite{jin_Raedt:2013,rossini_fazio:2007,yuan_raedt:2006} and on the absence of coherence at the macroscopic scale~\cite{Braun_Strunz:2001,Stamp:2006}, one can expect that decoherence is facilitated when the bath size increases. The decoherence rate depends on the topology of the participating potential energy surfaces, which are complicated high-dimensional functions that depend on the whole molecular geometry. As a result, designing molecules without rapid decoherence is challenging. A better understanding of the decoherence mechanisms is required, in particular, for developing new devices or materials exploiting coherence~\cite{Scholes_Zhu:2017}.

Experimentally, charge migration has been observed, e.g., with photofragmentation~\cite{Calegari_Nisoli:2014}, sequential double ionization~\cite{Fleischer_Staudte:2011,Mansson_Calegari:2021}, high-harmonic generation (HHG) spectroscopy~\cite{Kraus_Worner:2015,Worner_Wenger:2017}, and attosecond transient absorption spectroscopy (ATAS)~\cite{Goulielmakis_Krausz:2010, Matselyukh_Worner:2022}. HHG offers spatial and time resolution of the hole dynamics, but only for the first few femtoseconds, because of a limited number of generated harmonics, thus missing the important effect of the nuclear motion on a longer time scale. For this reason, ATAS has emerged as a powerful technique for tracking hole migration with time and spatial resolution for tens of femtoseconds by exploiting element-specific core-to-valence transitions~\cite{Golubev_Kuleff:2021}. Going beyond zero-temperature and vacuum conditions, a recent work showed that resonant Raman spectroscopy can be used to disentangle the contributions from the solvent and individual vibrational modes to the overall decoherence~\cite{Gustin_Franco:2023}.

In principle, the evaluation of the electronic coherence requires computing the coupled electron-nuclear dynamics by solving the time-dependent Schr\"odinger equation for the molecular system. Due to the exponential scaling of the exact solution, approximations are necessary to treat molecules with more than just a few degrees of freedom.
One possible method is the popular surface-hopping algorithm~\cite{Hammes-Schiffer_Tully:1994}, provided that the swarm of trajectories has a defined nuclear density. This is typically achieved by adding a Gaussian density centered at each nuclear configuration~\cite{Ha_Min:2018}. In addition, various ``decoherence corrections''~\cite{Bittner_Rossky:1995,Hammes-Schiffer_Fang:1999,Jasper_Truhlar:2002,Granucci_Persico:2010,Subotnik_Bellonzi:2016} have been developed to ensure consistency between the repartition of the trajectories among the electronic states and the electronic populations derived from the electronic wavefunction. However, a multi-trajectory approach already results in a significant computational overhead. Instead, several authors used a short-time approximation to estimate the initial coherence decay~\cite{Bittner_Rossky:1995,Schwartz_Rossky:1996, Gu_Franco:2018}. Unfortunately, the constraint of a Gaussian decay can become inaccurate after a few femtoseconds and certainly cannot capture the recoherence effect. Our approach, described in the Theory section, offers a compromise by propagating a single semiclassical nuclear wavepacket on each electronic surface using the single-Hessian thawed Gaussian approximation (TGA)~\cite{Begusic_Vanicek:2019,Begusic_Vanicek:2022}, combined with accurate \textit{ab initio} evaluations of the potential energy information at every time step. Therefore, in contrast to the often used Brownian oscillator model~\cite{book_Mukamel:1995}, mode distortion and Duschinsky rotation~\cite{Duschinsky:1937} (intermode coupling) are included exactly, and anharmonicity is included at least partially. At sufficiently short times, when the nonadiabatic couplings can be neglected, this approach has been shown to have similar accuracy as the fully quantum multi-configurational time-dependent Hartree (MCTDH) method~\cite{Golubev_Vanicek:2020}.

Here, we investigate the influence of the molecular structure on decoherence and charge migration by reducing or extending the carbon backbone of the but-3-ynal molecule, while conserving the same chemical properties. Surprisingly, we find that increasing the size and flexibility of a molecule can prolong both the electronic coherence and charge migration. Moreover, we explain the enhanced coherence in pent-4-ynal by showing that extending the molecular backbone suppresses decoherence induced by specific internal vibrations. This contrasts with a previous study~\cite{Gustin_Franco:2023}, where the increased coherence time in the (larger) nucleotide thymidine monophosphate compared to the (smaller) nucleobase thymine was explained by a chemical substitution that weakened the interaction with the solvent. Furthermore, prolonged coherence alone does not ensure long-lasting electron motion, as several additional properties are necessary for the coherence to be observable. Therefore, we analyze the effect of the carbon chain length on the ionization spectrum, on the duration and frequency of the electronic oscillation, and on the localization of the migrating charge, which are the key features needed for long-lasting charge migration after photoionization ~\cite{Scheidegger_Vanicek:2023}. Exploiting the semiclassical description of the nuclear motion, we find not only which vibrational modes are responsible for the suppression of the charge oscillation, but also through which mechanisms.

\section*{Theory}\label{sec:theory}
Before ionization, the molecule is assumed to be in its ground electronic state. Within the sudden ionization approximation, an electron is removed from this configuration, which is then projected onto the cationic subspace of the system, giving the amplitude $a_{I}$ of each populated cationic eigenstate $I$. To find the nuclear part of the initial state, we approximated the ground neutral electronic surface by a harmonic expansion at the equilibrium geometry. In the zero-temperature limit, the initial state is a Gaussian wavepacket, which, upon electronic excitation, is vertically duplicated in each populated cationic state $I$ and weighted by the respective coefficient $a_{I}$. Each wavepacket,
\begin{equation}
\label{eq:TGA_WP}
\begin{split}
	\chi_{I}(\mathbf{R},t)=a_{I}\exp\bigg\{
		\frac{i}{\hbar} \bigg[
		\frac{1}{2}(\mathbf{R}-\mathbf{R}^{I}_{t})^{T} \cdot \mathbf{A}^{I}_{t} \cdot (\mathbf{R}-\mathbf{R}^{I}_{t}) \\
		+ (\mathbf{P}^{I}_{t})^{T} \cdot (\mathbf{R}-\mathbf{R}^{I}_{t}) + \gamma^{I}_{t}
		\bigg]
	\bigg\},
\end{split}
\end{equation}
is uniquely defined by four time-dependent parameters: $\mathbf{R}^{I}_{t}$ and $\mathbf{P}^{I}_{t}$ are the position and momentum of the center of the wavepacket in the mass- and frequency-scaled coordinates of the neutral harmonic potential, $\mathbf{A}^{I}_t$ is a complex symmetric matrix with a positive definite imaginary part controlling the width of the Gaussian, and $\gamma^{I}_{t}$ is a complex number whose real part adds a dynamical phase, while its imaginary part guarantees normalization at all times.

In the original TGA, the ansatz (\ref{eq:TGA_WP}) and a local second-order Taylor expansion of the potential at the center of the Gaussian wavepacket are inserted into the time-dependent Schr\"odinger equation~\cite{Heller:1975}. This leads to a system of ordinary differential equations describing the time evolution of the four parameters~\cite{Heller:1975,Begusic_Vanicek:2020}. In this paper, we employed the single-Hessian variant~\cite{Begusic_Vanicek:2019,Begusic_Vanicek:2022} of the method, where a single reference Hessian (per electronic state) is used throughout the propagation, which results not only in higher efficiency but also in exact conservation of the effective energy and symplectic structure~\cite{Vanicek:2023}.

From the Born-Huang representation~\cite{book_Born_Huang:1954} of the molecular wavefunction, the expectation value of a molecular operator $\hat{O}(\mathbf{r},\mathbf{R})$ can be written as
\begin{equation}
\label{eq:exp_value}
	\langle\hat{O}\rangle(t)=\sum_{I,J}\int\chi_{I}^{*}(\mathbf{R},t)O_{IJ}(\mathbf{R})\chi_{J}(\mathbf{R},t)d\mathbf{R},
\end{equation}
where $O_{IJ}(\mathbf{R}):=\int\Phi_{I}^{*}(\mathbf{r},\mathbf{R})\hat{O}(\mathbf{r},\mathbf{R})\Phi_{J}(\mathbf{r},\mathbf{R})d\mathbf{r}$. If both the operator $\hat{O}(\mathbf{r},\mathbf{R})$, and the electronic states $\Phi_{I}(\mathbf{r},\mathbf{R})$ and $\Phi_{J}(\mathbf{r},\mathbf{R})$ depend on the nuclear coordinates $\mathbf{R}$ only weakly, Eq.~(\ref{eq:exp_value}) reduces to
\begin{equation}
\label{eq:exp_value_approx}
\langle \hat{O} \rangle(t) \approx \sum_{I,J} O_{IJ} \chi_{IJ}(t),
\end{equation}
where all the time dependence of the expectation value is contained in the electronic coherence
\begin{equation}
\label{eq:coherence_gwp}
	\chi_{IJ}(t) = \int \chi_{I}(\mathbf{R},t)^{*} \chi_{J}(\mathbf{R},t) d\mathbf{R}.
\end{equation}
The formula for the overlap of two Gaussian wavepackets of the form (\ref{eq:TGA_WP}) can be found in Ref.~\onlinecite{Begusic_Vanicek:2020}.
\begin{figure*}[t]
\centering
\includegraphics[]{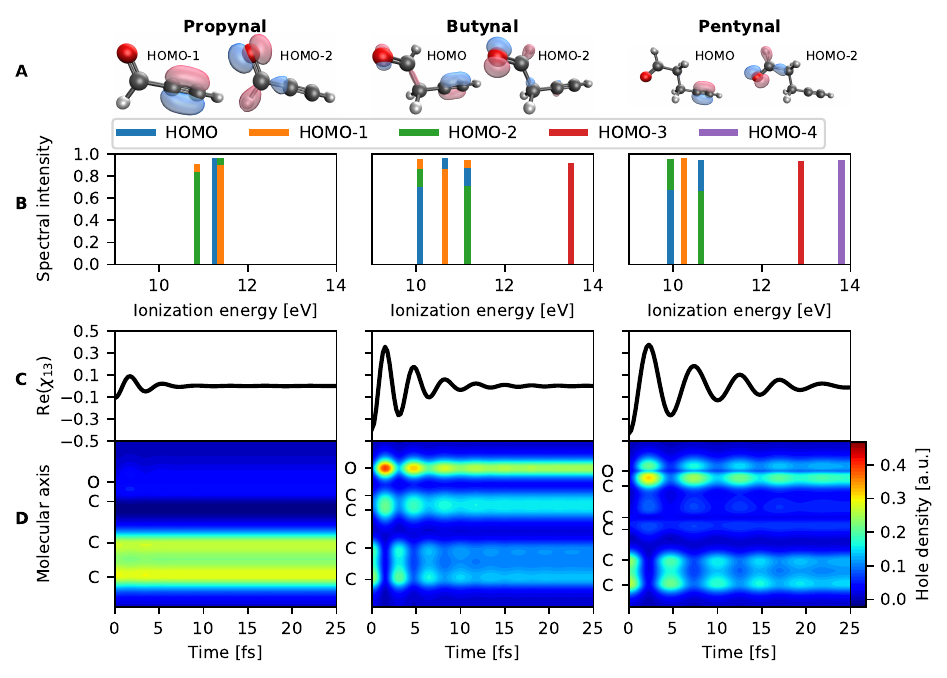}
\caption{Ionization spectrum and charge migration triggered by ionizing the propynal (left), butynal (middle), and pentynal (right) molecules. (A) Hartree-Fock molecular orbitals involved in the hole-mixing. (B) First cationic states in the energy range of 9 to 14 eV. (C) Time evolution of the electronic coherence between the first and third cationic states created by the ionization. (D) Charge migration projected along the molecular axis passing through the carbon triple bond.}
\label{fig:prop_but_pent_CM}
\end{figure*}
Because decoherence is determined by short-time dynamics, our calculations use the ``vertical'' Hessian~\cite{Begusic_Vanicek:2019} at the Franck-Condon point as the single reference Hessian per surface. Such a method is exact in any global harmonic potentials. Alternatively, one may use the same, “initial” Hessian, i.e., the Hessian at the minimum of the ground-state surface of the neutral molecule, as a reference Hessian in both excited states, and obtain dynamics in which the widths of the propagated Gaussians remain fixed~\cite{Begusic_Vanicek:2019} as in the frozen Gaussian approximation~\cite{Heller:1981}. Although the initial single-Hessian method is no longer exact in distorted and rotated harmonic potentials, it is generally almost as accurate as the vertical single-Hessian TGA. This is because the position and momentum of the Gaussian still follow the exact classical trajectory on each anharmonic surface, thereby capturing most of the anharmonicity, Duschinsky-rotation, and mode-distortion effects. Moreover, this method permits simplifying Eq.~(\ref{eq:coherence_gwp}) for the electronic coherence, yielding an intuitive expression~\cite{Golubev_Vanicek:2020}
\begin{equation}\label{eq:coherence_anl}
\chi_{IJ} (t) = a^{*}_{I} a_{J}e^{-d(t)^2/4\hbar} e^{iS(t)/\hbar},
\end{equation}
where $S(t)$ is the ``classical action'' (see Refs.~\onlinecite{Golubev_Vanicek:2020,Zambrano_Almeida:2011} for a precise definition) and
\begin{equation}\label{eq:squared_PS_dist}
d(t)^{2} = |\mathbf{R}_{t}^{I}-\mathbf{R}_{t}^{J}|^2 + |\mathbf{P}_{t}^{I}-\mathbf{P}_{t}^{J}|^2
\end{equation}
is the squared phase-space distance between the endpoints of the two trajectories propagated on the $I$th and $J$th surfaces.

The form of Eq.~(\ref{eq:coherence_anl}) further simplifies semiclassical interpretation of decoherence~\cite{Golubev_Vanicek:2020}. This strategy was applied in Ref.~\onlinecite{Matselyukh_Worner:2022}, where a modified expression for the coherence decay $e^{-d(t)^{2}}$ was used  by replacing $\mathbf{R}_{t}^{I}$ and $\mathbf{P}_{t}^{I}$ in Eq.~(\ref{eq:squared_PS_dist}) with the expectation values of position and momentum in the molecular wavefunction propagated with the MCTDH method. This approach successfully explained the experimentally observed decoherence and recoherence of the silane molecule by the time dependence of a phase-space overlap of the two wavepackets.

Without any further approximation, the coherence decay in Eq.~(\ref{eq:coherence_anl}) can be expressed as a product of the individual position and momentum contributions from each nuclear mode $j$,
\begin{equation}\label{eq:prod_indiv_contr}
e^{-d(t)^2/4\hbar} = \prod_{j}e^{-(R_{t,j}^{J}-R_{t,j}^{I})^2/4\hbar} e^{-(P_{t,j}^{J}-P_{t,j}^{I})^2/4\hbar}.
\end{equation}
In this form, the semiclassical description of nuclear dynamics offers a powerful tool for disentangling the overall decoherence induced by the nuclear motion.
In our simulations based on Eq.~(\ref{eq:coherence_gwp}), the wavepacket widths evolve over time, allowing a more accurate evaluation of the electronic energies and consequently of the oscillation frequency of the migrating charge. In contrast, the influence of the evolution of the width on the coherence decay is expected to be minimal, preserving the validity of Eqs.~(\ref{eq:coherence_anl}) and (\ref{eq:prod_indiv_contr}).

\section*{Results}
We recently showed~\cite{Scheidegger_Golubev:2022} that the but-3-ynal molecule (hereafter referred to as butynal) presented charge migration lasting for about 10 fs after ionization out of the HOMO. Conserving the aldehyde and alkyne group at the ends of the structure, we shortened or lengthened the carbon skeleton. We thus obtained propynal and pent-4-ynal (pentynal), illustrated in Fig.~\ref{fig:prop_but_pent_CM}(A), which also shows the molecular Hartree--Fock orbitals involved in the charge migration.
The HOMO-2 is localized at the aldehyde group and has a $\pi$-antibonding shape in all three molecules. The HOMO (HOMO-1 in propynal due to a swap of two energy levels) has the shape of a $\pi$-bonding orbital located at the carbon-carbon triple bond. The different energy ordering of the molecular orbitals in propynal might be caused by the close proximity of the two chemical groups, which creates a mesomeric effect delocalizing the electron of the alkyne to the aldehyde group.

\begin{figure}
\centering
\includegraphics[]{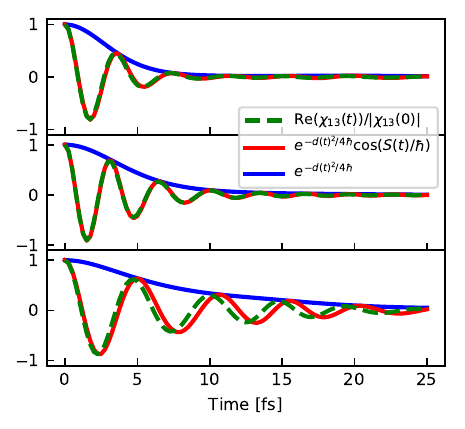} \caption{Semiclassical analysis of the electronic coherence in propynal (top), butynal (middle) and pentynal (bottom). Coherences computed with
the numerical Eq.~(\ref{eq:coherence_gwp}) (dash-dotted green line) and quasi-analytical Eq.~(\ref{eq:coherence_anl}) (solid red line) are compared. The overall decay of the coherence (solid blue line) depends on the phase-space distance $d(t)$.}
\label{fig:prop_pent_SC_anl}
\end{figure}
Figure~\ref{fig:prop_but_pent_CM}(B) shows that the ionization spectra of the three molecules are similar. In all cases, a hole-mixing structure~\cite{Breidbach_Cederbaum:2003} is present between the first and third cationic states, which are primarily composed of the previously mentioned ionized orbitals. Propynal and pentynal have a planar symmetry and belong to the $C_{s}$ point group. The mixed cationic states as well as the involved orbitals have $A^\prime$ symmetry, whereas the second cationic state has the $A^{\prime\prime}$ symmetry. Experimentally, selective excitation of only the first and third cationic states can be achieved by orienting the molecule with respect to the polarization of the laser pulse. In contrast, butynal does not have any symmetry, as the OCCC dihedral angle differs by $30^{\circ}$ from the less favorable symmetric conformer (see SI Appendix, Fig.~S1). This induces a slight hole-mixing between the second cationic eigenstate and the adjacent ones. However, the charge migration dynamics arising from the possible presence of the additional electronic state in the wavepacket is negligible since the major contribution to the second state is coming from the orthogonal HOMO-1 orbital.

The hole-mixing intensity, crucial prerequisite for charge migration~\cite{Breidbach_Cederbaum:2003,Scheidegger_Vanicek:2023}, increases with the number of carbon atoms (see Figs.~S2 and S3 in the SI Appendix for the ionization spectra of hexynal and heptynal). This is particularly important, since the coherence (\ref{eq:coherence_anl}) is maximized when the electronic populations are equal, i.e., $|a_{I}|^2=|a_{J}|^2=1/2$, which happens when the hole-mixing is maximal and the initial state involves only two 1h configurations~\cite{Breidbach_Cederbaum:2003}.

In Fig.~\ref{fig:prop_but_pent_CM}(C), the real part of the electronic coherence is shown. In propynal, the coherence oscillation has a very small amplitude because of the weak hole-mixing, whereas pentynal has a more pronounced and more persistent coherence oscillation than butynal. This result is remarkable, since adding a carbon atom makes the molecule more flexible and increases the number of vibrational modes, which would be expected to accelerate the decoherence induced by the nuclear motion.
One may object that the asymmetry of butynal and different initial electronic populations prevent a meaningful comparison. To eliminate these effects and make the comparison more direct, we show, in Fig.~S4 of the SI Appendix, the absolute value of the normalized coherence for all three molecules, as well as for the $C_s$ conformer of butynal. We thus demonstrate that the enhancement of the electronic coherence lifetime does not rely on specific initial electronic populations or molecular conformation.

The charge migration analysis in Fig.~\ref{fig:prop_but_pent_CM}(D) shows the hole dynamics after ionization out of the HOMO (HOMO-1 for propynal). Due to the very weak hole-mixing in propynal, the hole remains strongly localized at the triple bond, and no oscillation can be observed. Charge migration is more pronounced in pentynal than butynal due to the disappearance of the hole density at the C-C bond close to the oxygen. This suggests that experimental observation of the hole dynamics with time and spatial resolution might be easier in pentynal.

\begin{figure}
\centering
   \includegraphics[]{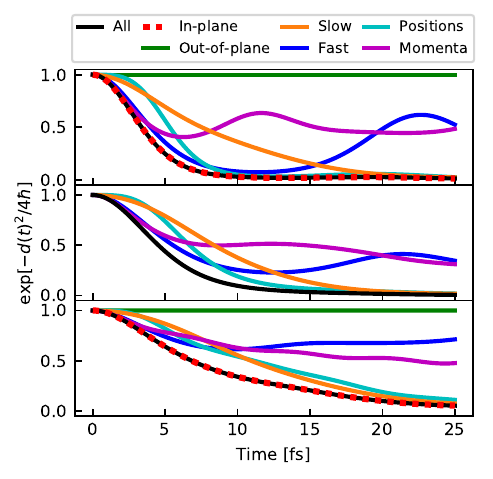} \caption{Decay of the electronic coherence in propynal (top), butynal (middle) and pentynal (bottom) when including all vibrational modes, only the in-plane $A^{\prime}$ modes, only the out-of-plane $A^{\prime\prime}$ modes, only the slow modes ($\le$1100 $\mathrm{cm}^{-1}$), only the fast modes ($>$1100 $\mathrm{cm}^{-1}$), and when including all modes but separating their position and momentum contributions.}
\label{fig:prop_pent_sym_NM}
\end{figure}
Figure~\ref{fig:prop_pent_SC_anl} compares the quasi-analytical expression (\ref{eq:coherence_anl}) for the electronic coherence with the coherence computed as the overlap (\ref{eq:coherence_gwp}) of the Gaussian wavepackets. The excellent agreement validates the more approximate quasi-analytical expression in all three molecules. In propynal and butynal, the two curves are indistinguishable, whereas in pentynal a slight dephasing appears due to different width evolution of the two wavepackets. This difference, however, has a negligible effect on the decay of the coherence, which is the most important feature for our purpose.

\begin{figure*}[t]
\centering
\includegraphics[]{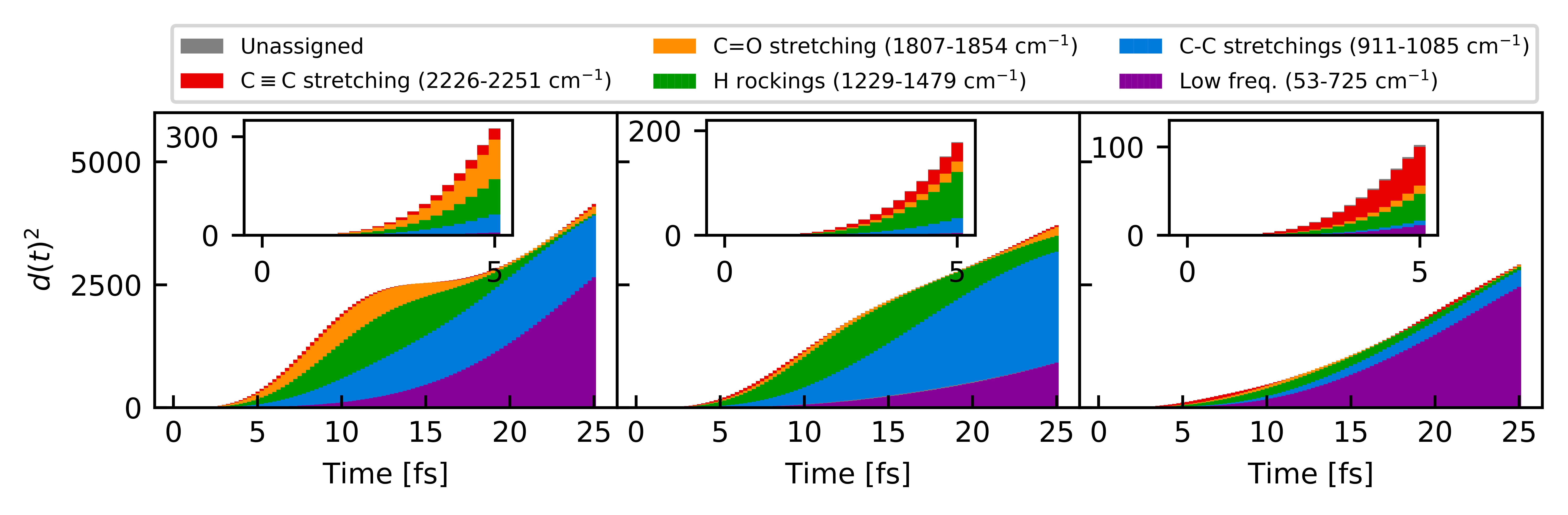} \caption{Decomposition of the squared phase-space distance into contributions from individual normal mode categorized by types of vibrational motion in propynal (left), butynal (middle) and pentynal (right). The low-frequency vibrations involve the whole molecular structure and cannot be assigned to a standard type of motion. Several modes with negligible contributions have not been assigned (gray).}
\label{fig:prop_but_pent_NM_weight_type_of_motion}
\end{figure*}
Using Eq.~(\ref{eq:prod_indiv_contr}), we show in Fig.~\ref{fig:prop_pent_sym_NM} that the overall decay of the electronic coherence is caused by an interplay of the momentum and position separations of the wavepackets evolving in different electronic states. However, the contributions of the two mechanisms vary at different times. During the first ~2-3 fs, the decoherence is primarily due to the increase in the momentum separation, as predicted by the short time approximation model of Schwartz \textit{et al}~\cite{Schwartz_Rossky:1996} and, more recently, in Ref.~\onlinecite{Vacher_Malhado:2017}. This effect appears as the decay of the density overlap in the momentum representation and as dephasing in the position representation. After the first few femtoseconds, the position separation starts to dominate, whereas the momentum decoherence tends to stabilize.

We also classified nuclear vibrations as slow or fast, using a cutoff value of 1100 $\mathrm{cm}^{-1}$. The fast modes induce the initial decoherence but retain a high degree of coherence on the longer time scale. In propynal, this leads to a pronounced recoherence after 15 fs if only the fast modes are included. In contrast, slow vibrations play a minor role initially but ultimately suppress electronic coherence entirely. Our observations are consistent with those reported for thymine in Ref.~\onlinecite{Gustin_Franco:2023}, where the (fast) molecular vibrations induced the initial coherence decay, and the (slow)  solvent modes caused the eventual complete loss of coherence.

Dividing the normal modes between the totally symmetric in-plane $A^{\prime}$ vibrations and the out-of plane $A^{\prime\prime}$ vibrations, we observe that only the normal modes that preserve the symmetry of the molecule are responsible for decoherence (see Fig.~\ref{fig:prop_pent_sym_NM}). This distinction is not applicable to butynal, as all its vibrations belong to the same irreducible representation due to the lack of symmetry. The distribution between the two irreducible representations is $\Gamma_{vib} = 9A^{\prime}+3A^{\prime\prime}$ for propynal and $\Gamma_{vib} = 19A^{\prime}+11A^{\prime\prime}$ for pentynal. The absence of contribution from the out-of-plane vibrations has been recently observed also by Vester, Despr\'e, and Kuleff who applied the MCTDH method to the vibronic-coupling Hamiltonian model of the propynamide molecule ($C_{s}$ point group)~\cite{Vester_Kuleff:2023}. To identify the normal modes responsible for decoherence, their approach requires multiple simulations, in which the dimension of the system is gradually decreased until long electronic coherence appears. In contrast, our semiclassical approach neglects nonadiabatic effects but finds the decoherence-inducing modes at a much lower cost from a single simulation using accurate, full-dimensional \textit{ab initio} potential energy surfaces.

In Fig.~\ref{fig:prop_but_pent_NM_weight_type_of_motion}, we show the individual contribution of each totally symmetric normal mode to the squared phase-space distance (\ref{eq:squared_PS_dist}) according to its type of motion and vibrational frequency. We remind the reader that in butynal ($C_{1}$ point group),  all modes are associated with the totally symmetric $A$ representation. In Fig.~\ref{fig:prop_but_pent_NM_weight_type_of_motion}, we chose to display $d(t)^{2}$, the squared phase-space distance,  to benefit from the additivity over the normal modes, but we emphasize that the coherence decay $e^{-d(t)^{2}/4\hbar}$ is exponentially faster (see Fig.~\ref{fig:prop_pent_sym_NM}). In all three molecules, the vibrational frequencies range approximately from 50 to 3500 $\mathrm{cm}^{-1}$. Decoherence in the first 5 fs is induced by the fast C$\equiv$C and C$=$O stretching modes, and various H rocking vibrations. On the longer time scale the decoherence is dominated by the C-C stretching and other slow vibrational modes involving the whole molecule. In contrast to Refs.~\onlinecite{Arnold_Santra:2017,Vester_Kuleff:2023}, we find that the hydrogen vibrations contribute substantially to decoherence. Comparing the left and middle panels, one can conclude that butynal has a longer electronic coherence than propynal due to the absence of the 1808 $\mathrm{cm}^{-1}$ vibration, corresponding to the C=O stretching. Similarly, the improvement of pentynal over butynal is caused by a significant reduction in the contribution of the 1302 $\mathrm{cm}^{-1}$ mode, associated with one of the rocking motions of hydrogen atoms.
%

Since the $C_{s}$ point group is Abelian, all normal modes are nondegenerate.  At the Franck-Condon point, each electronic surface has a zero gradient along each of the asymmetric (out-of-plane) $A^{\prime\prime}$ modes~\cite{Domcke_Koppel:2004,Neville_Schuurman:2022}, which results in a slow wavepacket motion and slow decoherence along these modes. In contrast, the gradients along symmetric (in-plane) $A^{\prime}$ modes can be non-zero, with potentially opposite signs on the two different surfaces, pushing the associated wavepackets in different directions and resulting in a quick suppression of the electronic coherence. However, this ``gradient directed" decoherence is not the only possible mechanism within the TGA. A decrease in the overlap of the wavepackets can also occur if their widths evolve differently, but our results suggest that this effect is negligible in the molecules analyzed here.

While our propagation method is efficient, it is more expensive than other commonly used~\cite{Gustin_Franco:2023} but but cruder approximations. To justify the necessity to use our semiclassical approach, in Fig.~S5(a) in the SI Appendix, we compared the coherence decay evaluated with the single-Hessian TGA to the results obtained with three increasingly severe harmonic approximations, which progressively ignore anharmonicity, mode mixing, and mode distortion. In addition, in Fig. S5(b) we evaluated the coherence decay with three increasingly crude short-time approximations: zeroth-order dephasing representation~\cite{Vanicek_Cohen:2016,Vacher_Robb:2015,Jenkins_Robb:2016}, anharmonic short-time approximation from Ref.~\onlinecite{Gu_Franco:2018}, and short-time approximation to the displaced harmonic oscillator model~\cite{Gu_Franco:2018}. The comparison reveals that all three short-time approximations greatly underestimate coherence time ($\sim 10$~fs vs.~25~fs in pentynal), whereas the harmonic methods overestimate it and produce non-physical recoherence. We conclude that relying on simple approximations may lead to missing particularly intriguing effects, such as the prolonged coherence reported in this paper, or significantly overestimating electronic coherence beyond the initial decay when neglecting anharmonicity.

Our semiclassical approach neglects nonadiabatic transitions, which can also induce decoherence~\cite{Fiete_Heller:2003}, and uses the adiabatic electronic states, whose energies along the nuclear trajectories are shown in Fig.~S6 of the SI Appendix. This figure supports the adiabatic approximation because the energy levels remain well separated before the coherence is suppressed by pure dephasing. To justify the use of adiabatic states, which carry some entanglement between electrons and nuclei, for evaluating electronic coherences, in the SI Appendix we demonstrate that the adiabatic electronic states depend on nuclear coordinates only weakly. In particular, in Fig. S7, we show that the overlap between the initial and propagated electronic wavefunctions remains very high (above ~0.85) throughout the charge migration process.

To assess the importance of nonadiabatic dynamics quantitatively, we performed fewest-switches surface-hopping~\cite{Hammes-Schiffer_Tully:1994} simulations (see Figs.~S8 and S9). We found that, despite a slight population transfer to the second state in propynal and pentynal, nonadiabatic effects do not play a significant role in the evolution of the electronic coherences and thus do not influence the corresponding charge migration.

\section*{Conclusion}
\label{sec:conclusions}
We studied the effect of molecular structure on  charge migration triggered by valence ionization. Starting from butynal, a molecule already shown to exhibit long-lasting charge migration, we shortened or lengthened the carbon backbone to obtain the propynal and pentynal compounds. To our surprise, we observed that increasing the number of atoms and flexibility of the molecule, while conserving its chemical properties, led to longer charge migration. The cause was demonstrated to be the disappearance of the decoherence contributions of several normal modes and an increase of the hole-mixing.
It is unlikely that increasing the size of a molecule will consistently prolong electronic coherence, since coherence is inherently a quantum effect. Nevertheless, our observations and explanation of this phenomenon represent an important step forward in the understanding of general principles underlying electron-nuclear correlations and could be used for intelligent design of molecules with long-lasting electronic coherence and charge migration.
Using a semiclassical description of the nuclear dynamics allowed us to analyze the individual contributions of all normal modes to the electronic decoherence without additional computational cost. We confirmed that in molecules with an Abelian point group (here $C_{s}$), only the totally symmetric modes induce decoherence, while the contributions of the remaining ones are completely negligible. The decoherence happens as the nuclear wavepackets evolving in distinct electronic states move to different regions of the phase space. In the first few femtoseconds, the nuclear decoherence is induced by the momentum separation (dephasing), until the position contribution (the decay of nuclear density overlap) becomes more important and results in a complete fading of the charge migration.

\section{Materials and Methods}
\subsection*{Semiclassical dynamics}
The semiclassical dynamics method that we used is described in the Theory section. The Gaussian wavepackets were propagated with a time step of 0.25 fs for 100 steps using the semiclassical generalization of the Verlet algorithm~\cite{Vanicek:2023}.

\subsection*{Electronic structure}
All molecular geometries were optimized in the neutral ground state at the $\omega$B97XD/6-311++G(d,p) level of theory using the Gaussian 16 package~\cite{Frisch_Fox:2016}. Ionization spectra were computed with the algebraic diagrammatic construction (ADC) scheme~\cite{Schirmer:1982,Schirmer_Walter:1983} using the Hartree--Fock orbitals obtained with the GAMESS-UK 7.0 package~\cite{Guest_Kendrick:2005} and the double-zeta plus polarization (DZP)~\cite{Canal_Jorge:2005} basis set. The energies, gradients, and Hessians for the nuclear wavepackets propagation were computed with the equation-of-motion coupled-cluster for ionization potential (EOM-CC-IP)~\cite{Stanton_Bartlett:1993,Gour_Piecuch:2006,Kamiya_Hirata:2006} and the DZP basis set using the Q-Chem package~\cite{Shao_Head-Gordon:2015}. The Cartesian coordinates of the optimized geometries are shown in SI Appendix, Tables S1-S6.

\subsection*{Data, Materials, and Software Availability}
All study data are included in the article and/or SI Appendix.

\section*{Acknowledgments}
A.S. and J.J.L.V. acknowledge financial support from the Swiss National Science Foundation through the National Center of Competence in Research MUST (Molecular Ultrafast Science and Technology) and from the European Research Council (ERC) under the European Union's Horizon 2020 research and innovation program (Grant Agreement No. 683069--MOLEQULE). N.V.G. acknowledges support from the U.S. Department of Energy (DOE), Office of Science, Basic Energy Sciences (BES) under Award \#DE-SC0024182.

\bibliography{can_increasing_size_and_flexibility_of_molecule_reduce_decoherence}

\end{document}


\title{Supporting Information for: Can increasing the size and flexibility of a molecule reduce decoherence and prolong charge migration?}
\author{Alan Scheidegger}
\email{alan.scheidegger@epfl.ch}
\affiliation{Laboratory of Theoretical Physical Chemistry, Institut des Sciences et
Ing\'enierie Chimiques, Ecole Polytechnique F\'ed\'erale de Lausanne (EPFL),
CH-1015, Lausanne, Switzerland}

\author{Nikolay V. Golubev}
\email{ngolubev@arizona.edu}
\affiliation{Department of Physics, University of Arizona, 1118 E. Fourth Street, Tucson, AZ 85721}

\author{Ji\v{r}\'{\i} J. L. Van\'{\i}\v{c}ek}
\email{jiri.vanicek@epfl.ch}
\affiliation{Laboratory of Theoretical Physical Chemistry, Institut des Sciences et
Ing\'enierie Chimiques, Ecole Polytechnique F\'ed\'erale de Lausanne (EPFL),
CH-1015, Lausanne, Switzerland}

\date{\today}

\maketitle

\section*{Geometries and symmetry}

\begin{figure}[H]
\centering
\includegraphics[width=\textwidth]{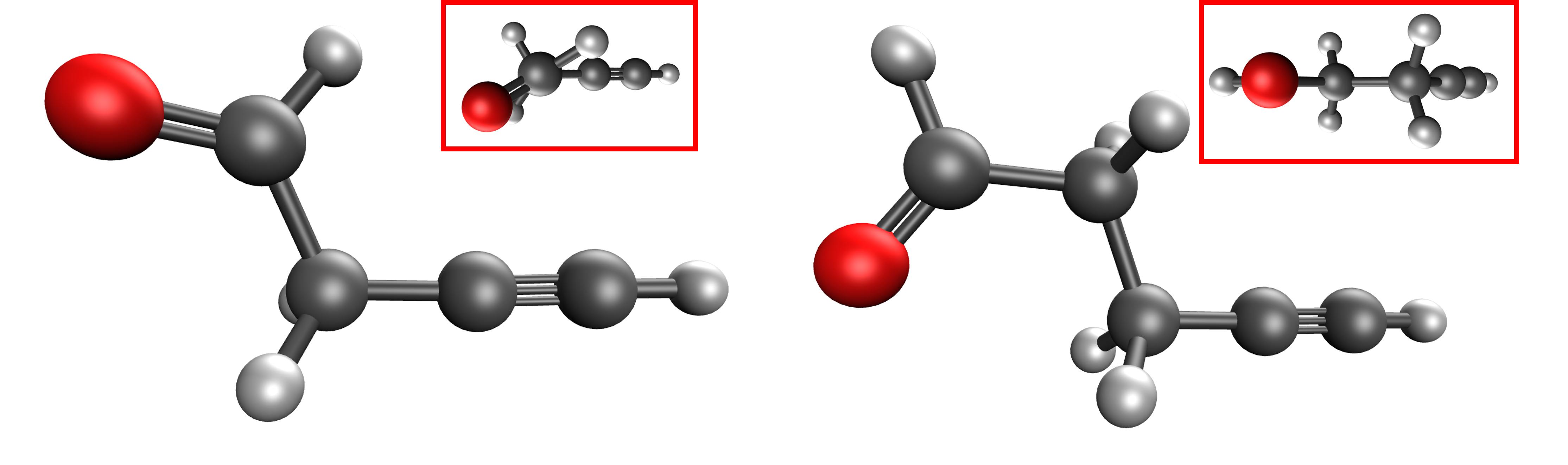}
\caption{Molecular geometries of butynal (left) and pentynal (right). Pentynal has a planar symmetry and belongs to the $C_{s}$ point group, whereas butynal has the aldehyde group rotated by $30\degree$ compared to the planar geometry.}
\end{figure}
All geometries have been optimized at the $\omega$B97XD/6-311++G(d,p) level.

\begin{table}[H]
\centering
\caption{Cartesian coordinates of propynal.}
\resizebox{0.29\columnwidth}{!}{%
\begin{tabular}{lrrr}
Atom & X & Y & Z \\
\midrule
C & -1.862526  & -0.133679  &  0.000000\\
C & -0.682909  &  0.090546  &  0.000004\\
C &  0.732185  &  0.410499  &  0.000000\\
O &  1.603232  & -0.418473  & -0.000001\\
H & -2.905643  & -0.347742  & -0.000014\\
H &  0.959289  &  1.491328  & -0.000007\\
\bottomrule
\end{tabular}
}
\end{table}

\begin{table}[H]
\centering
\caption{Cartesian coordinates of butynal.}
\resizebox{0.29\columnwidth}{!}{%
\begin{tabular}{lrrr}
Atom & X & Y & Z \\
\midrule
 C    &     1.089018   &  -0.020648   &   0.053526\\
 C    &     2.286528   &   0.039022   &   0.069112\\
 C    &     3.741575   &   0.082109   &   0.128505\\
 C    &     4.310740   &  -0.831556   &   1.206952\\
 O    &     5.328048   &  -0.594689   &   1.793293\\
 H    &     0.026374   &  -0.064751   &   0.031336\\
 H    &     4.110561   &   1.096314   &   0.297635\\
 H    &     4.159987   &  -0.259314   &  -0.827222\\
 H    &     3.727169   &  -1.752967   &   1.404492\\
\bottomrule
\end{tabular}
}
\end{table}

\begin{table}[H]
\centering
\caption{Cartesian coordinates of the symmetric conformer of butynal.}
\resizebox{0.29\columnwidth}{!}{%
\begin{tabular}{lrrr}
Atom & X & Y & Z \\
\midrule
C  &  -2.422237  & -0.404892  &  0.000000\\
C  &  -1.337408  &  0.106368  &  0.000000\\
C  &  -0.000000  &  0.682293  &  0.000000\\
C  &   1.113440  & -0.362574  &  0.000000\\
O  &   2.273153  & -0.062942  &  0.000000\\
H  &  -3.390204  & -0.846429  &  0.000000\\
H  &   0.148436  &  1.322041  &  0.876236\\
H  &   0.148436  &  1.322041  & -0.876236\\
H  &   0.785335  & -1.421286  &  0.000000\\
\bottomrule
\end{tabular}
}
\end{table}

\begin{table}[H]
\centering
\caption{Cartesian coordinates of pentynal.}
\resizebox{0.29\columnwidth}{!}{%
\begin{tabular}{lrrr}
Atom & X & Y & Z \\
\midrule
C  &   3.020901  &  0.074904  & -0.000289\\
C  &   1.854777  & -0.207971  &  0.000218\\
C  &   0.429999  & -0.524857  &  0.000273\\
C  &  -0.437276  &  0.734654  &  0.000052\\
C  &  -1.913872  &  0.445780  &  0.000031\\
O  &  -2.394628  & -0.655918  & -0.000254\\
H  &   4.057159  &  0.314276  & -0.000474\\
H  &   0.189870  & -1.136984  &  0.873894\\
H  &   0.189832  & -1.137299  & -0.873112\\
H  &  -0.220485  &  1.363715  & -0.871927\\
H  &  -0.220438  &  1.364131  &  0.871700\\
H  &  -2.566089  &  1.344445  &  0.000242\\
\bottomrule
\end{tabular}
}
\end{table}

\begin{table}[H]
\centering
\caption{Cartesian coordinates of hexynal.}
\resizebox{0.29\columnwidth}{!}{%
\begin{tabular}{lrrr}
Atom & X & Y & Z \\
\midrule
C    &   3.463575  & -0.709323  & -0.000091\\
C    &   2.481058  & -0.020634  &  0.000079\\
C    &   1.268971  &  0.793389  &  0.000187\\
C    &  -0.011802  & -0.051596  & -0.000083\\
C    &  -1.258769  &  0.820439  & -0.000049\\
C    &  -2.555206  &  0.054478  & -0.000207\\
O    &  -2.644503  & -1.143771  &  0.000206\\
H    &   4.336437  & -1.316835  & -0.000510\\
H    &   1.283302  &  1.449282  & -0.877537\\
H    &   1.283220  &  1.448896  &  0.878205\\
H    &  -0.019155  & -0.707698  &  0.873617\\
H    &  -0.018976  & -0.707377  & -0.874040\\
H    &  -1.282830  &  1.489302  & -0.871728\\
H    &  -1.282970  &  1.488955  &  0.871911\\
H    &  -3.469961  &  0.685123  & -0.000581\\
\bottomrule
\end{tabular}
}
\end{table}

\begin{table}[H]
\centering
\caption{Cartesian coordinates of heptynal.}
\resizebox{0.29\columnwidth}{!}{%
\begin{tabular}{lrrr}
Atom & X & Y & Z \\
\midrule
 C  &-0.677255 &  -0.222941  &  0.000488\\
 C  &-1.755373 &   0.851028  & -0.000190\\
 C  &-3.165456 &   0.324566  &  0.000389\\
 O  &-3.464156 &  -0.839878  & -0.000600\\
 H  &-0.812737 &  -0.867824  &  0.874850\\
 H  &-0.813011 &  -0.869212  & -0.872782\\
 H  &-1.661237 &   1.513296  & -0.872046\\
 H  &-1.660685 &   1.515244  &  0.870063\\
 H  &-3.956395 &   1.105095  &  0.001652\\
 C  & 0.729652 &   0.367530  & -0.000123\\
 C  & 1.811029 &  -0.721250  &  0.000276\\
 C  & 3.165114 &  -0.175025  & -0.000078\\
 C  & 4.268351 &   0.297589  & -0.000199\\
 H  & 0.869676 &   1.007987  &  0.877305\\
 H  & 0.869347 &   1.007005  & -0.878325\\
 H  & 1.686308 &  -1.365404  & -0.876855\\
 H  & 1.686398 &  -1.364633  &  0.877999\\
 H  & 5.249210 &   0.708480  & -0.000445\\
\bottomrule
\end{tabular}
}
\end{table}

\section*{Ionization spectra}

\begin{figure}[H]
\centering
\includegraphics[]{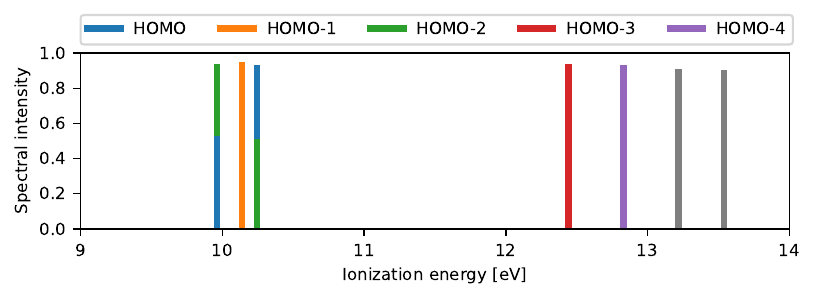}
\caption{Ionization spectrum of hexynal computed with the algebraic diagrammatic construction (ADC) scheme and the double-zeta plus polarization (DZP) basis set.}
\end{figure}

\begin{figure}[H]
\centering
\includegraphics[]{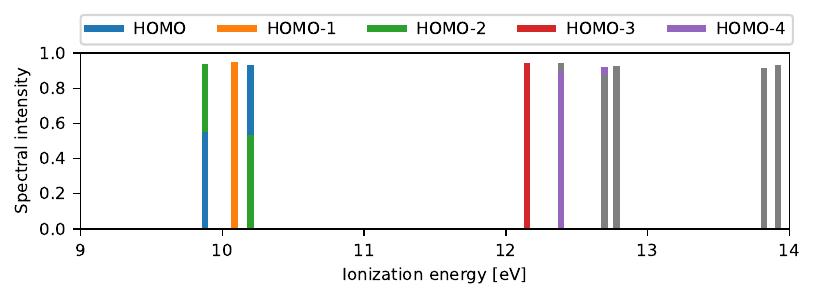}
\caption{Ionization spectrum of heptynal computed with the algebraic diagrammatic construction (ADC) scheme and the double-zeta plus polarization (DZP) basis set.}
\end{figure}

\section*{Normalized electronic coherence}
\begin{figure}[h]
\centering
\includegraphics[]{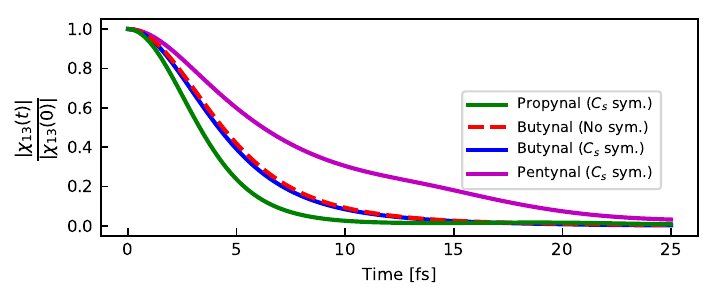}
\caption{Absolute value of the normalized electronic coherence for propynal, pentynal, and two conformers of butynal.}
\label{fig:butynal_Cs}
\end{figure}
Unlike propynal and pentynal, which have the planar $C_{s}$ symmetry, the most stable conformer of butynal has no symmetry. Additionally, the initial electronic populations after ionization are different for the three molecules because of different hole-mixing intensities. This has a direct effect on the amplitude of the electronic oscillation. The lack of symmetry of butynal and different initial electronic populations complicate the comparison of the coherence times when extending the carbon skeleton. In Fig.~\ref{fig:butynal_Cs}, we therefore show the normalized coherence $|\chi_{13}(t)|/ |\chi_{13}(0)|$, i.e. the coherence divided by its initial value, to exclude the effect of the electronic populations. Additionally, we included the normalized coherence following the excitation of the symmetric conformer of butynal. This coherence turns out to be essentially identical to the coherence of the most stable (asymmetric) conformer. We conclude that the prolongation of coherence with increasing size of the molecule remains even after removing the effects of the initial electronic populations and of different symmetry of butynal.

\section*{Comparison with harmonic and short-time approximations for coherence}
\begin{figure}[H]
\centering
\includegraphics[width=\textwidth]{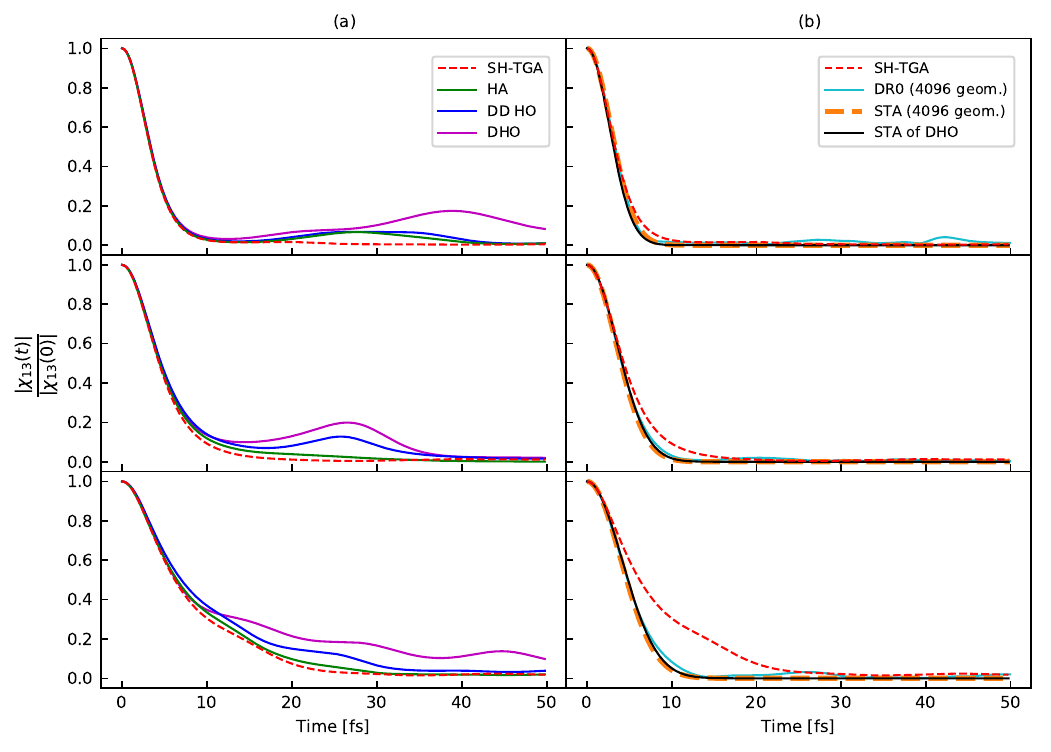}
\caption{Absolute value of the normalized electronic coherence of propynal (top), butynal (middle), and pentynal (bottom). Panel (a) compares  the single-Hessian TGA (SH-TGA) with three harmonic approximations: global harmonic approximation (HA), displaced and distorted harmonic oscillator (DD HO) model, and displaced harmonic oscillator (DHO) model. Panel (b) compares the SH-TGA with three short-time approximations: the zeroth-order dephasing representation (DR0), the anharmonic short-time approximation (STA), and the STA of the displaced harmonic oscillator model (STA of DHO).}
\label{fig:lower_approximations}
\end{figure}

In Fig.~\ref{fig:lower_approximations}, we compare the decay of the normalized electronic coherence evaluated with the single-Hessian TGA and with various more approximate methods. In panel (a), we show the results of three increasingly severe harmonic approximations. In the ``harmonic approximation'' (HA), a global harmonic approximation of the potential energy surfaces is constructed from a reference geometry (here the Franck-Condon point). This approximation captures the displacement of the minimum, mode distortion, and mode mixing (Duschinsky coupling), but neglects anharmonic effects. Without mode mixing, we get the ``displaced and distorted harmonic oscillator'' (DD HO) model, where the off-diagonal elements of the Hessians of the two excited-state surfaces are set to zero. The most approximate is the ``displaced harmonic oscillator'' (DHO) model, which also neglects mode distortion (i.e., changes in the force constants in the excited states), but still includes the horizontal displacement.

In panel (b), we compare our result with the results of three increasingly crude short-time approximations. The most accurate of the approximations that avoid propagation estimates the normalized coherence as
\begin{equation}
     \chi_{13}(t)/\chi_{13}(0) = \langle\exp[-it\Delta\varepsilon_{13}(\mathbf{R})/\hbar] \rangle_{|\chi(\mathbf{R},0)|^2},
\end{equation}
where $\Delta\varepsilon_{13}(\mathbf{R}):=E_{3}(\mathbf{R}) - E_{1}(\mathbf{R})$ is the energy gap between the third and first cationic states at position $\mathbf{R}$, and $\langle \cdots \rangle_{|\chi(\mathbf{R},0)|^2}$ denotes an average over $\mathbf{R}$ sampled from the initial nuclear density $|\chi(\mathbf{R},0)|^2$. This approximation, called ``zeroth-order dephasing representation'' (DR0) in Ref.~\onlinecite{Vanicek_Cohen:2016}, is the leading term in a path-integral expansion of coherence. It has been used by Robb and coworkers~\cite{Vacher_Robb:2015,Jenkins_Robb:2016} to evaluate decoherence in para-xylene and several derivatives of norbornadiene. Slav\'{\i}\v{c}ek and coworkers showed that this approximation was equivalent to the reflection principle of spectroscopy~\cite{Srsen_Slavicek:2021}. If the energy gap is normally distributed, the above expression reduces to another short-time approximation, in which the coherence depends only on the mean and variance of $\Delta\varepsilon_{13}$, i.e.,
\begin{equation}
    \chi_{13}(t)/\chi_{13}(0) = \exp[-it\langle\Delta\varepsilon_{13}\rangle/\hbar-t^{2}\mathrm{Var}(\Delta\varepsilon_{13})/2\hbar^{2}].
\end{equation}
This ``short-time approximation'' (STA), obtained e.g. in Ref.~\onlinecite{Gu_Franco:2018}, yields a Gaussian decay
$|\chi_{13}(t)/\chi_{13}(0)| = \exp(-t^2/2\tau_{13}^2)$    
of coherence with time scale
$    \tau_{13} = \hbar/\sqrt{\mathrm{Var}(\Delta\varepsilon_{13})}$.
In the DHO, the energy gap fluctuation $\mathrm{Var}(\Delta\varepsilon_{13})$ at zero temperature is
\begin{equation}
     \mathrm{Var}(\Delta\varepsilon_{13})= \frac{\hbar}{2}\sum_{j}m_{j}d_{j}^2\omega_{j}^3,
\end{equation}
where $m_{j}$ and $\omega_{j}$ are the mass and frequency associated with the nuclear coordinate $j$. The displacement $d_{j}$ of the minima of the two excited-state surfaces can be expressed from the gradients of the two surfaces at the Franck--Condon point as
\begin{equation}
    d_{j} = \frac{\nabla_{j}V_{1}-\nabla_{j}V_{3}}{m_{j}\omega_{j}^2}.
\end{equation}
Thus, the ``short-time approximation of the DHO'' (STA of DHO) can be evaluated analytically, whereas the first two short-time approximations require sampling the energy gap at $t=0$. Effects included in each approximation are summarized in Table~\ref{table:approx_effects}.

\begin{table}[h!]
\begin{center}
\captionof{table}{\textmd{Inclusion of various effects influencing electronic coherence by different approximations. Green = described well, red = not described at all, orange = described very roughly. Among the listed methods, only the single-Hessian TGA describes all effects.}\label{table:approx_effects}}
\begin{tabular}{ |c | c | c | c | c | c | }
\hline
& Decoherence & Recoherence & Mode distortion & Mode mixing & Anharmonicity \\
\hline
SH-TGA               & \cellcolor[HTML]{33ff3f}\hfil\ding{51} & \cellcolor[HTML]{33ff3f}\hfil\ding{51} & \cellcolor[HTML]{33ff3f}\hfil\ding{51} & \cellcolor[HTML]{33ff3f}\hfil\ding{51} & \cellcolor[HTML]{33ff3f}\hfil\ding{51} \\
\hline
HA                   & \cellcolor[HTML]{33ff3f}\hfil\ding{51} & \cellcolor[HTML]{33ff3f}\hfil\ding{51} & \cellcolor[HTML]{33ff3f}\hfil\ding{51} & \cellcolor[HTML]{33ff3f}\hfil\ding{51} & \cellcolor[HTML]{f11f25}\hfil\ding{53} \\
\hline
DDHO                 & \cellcolor[HTML]{33ff3f}\hfil\ding{51} & \cellcolor[HTML]{33ff3f}\hfil\ding{51} & \cellcolor[HTML]{33ff3f}\hfil\ding{51} & \cellcolor[HTML]{f11f25}\hfil\ding{53} & \cellcolor[HTML]{f11f25}\hfil\ding{53} \\
\hline
DHO                  & \cellcolor[HTML]{33ff3f}\hfil\ding{51} & \cellcolor[HTML]{33ff3f}\hfil\ding{51} & \cellcolor[HTML]{f11f25}\hfil\ding{53} & \cellcolor[HTML]{f11f25}\hfil\ding{53} & \cellcolor[HTML]{f11f25}\hfil\ding{53} \\
\hline
DR0   & \cellcolor[HTML]{33ff3f}\hfil\ding{51} & \cellcolor[HTML]{f1851f}\hfil\ding{51} & \cellcolor[HTML]{f1851f}\hfil\ding{51} & \cellcolor[HTML]{f1851f}\hfil\ding{51} & \cellcolor[HTML]{f1851f}\hfil\ding{51} \\
\hline
STA   & \cellcolor[HTML]{33ff3f}\hfil\ding{51} & \cellcolor[HTML]{f11f25}\hfil\ding{53} & \cellcolor[HTML]{f1851f}\hfil\ding{51} & \cellcolor[HTML]{f1851f}\hfil\ding{51} & \cellcolor[HTML]{f1851f}\hfil\ding{51} \\
\hline
STA of DHO   & \cellcolor[HTML]{33ff3f}\hfil\ding{51} & \cellcolor[HTML]{f11f25}\hfil\ding{53} & \cellcolor[HTML]{f11f25}\hfil\ding{53} & \cellcolor[HTML]{f11f25}\hfil\ding{53} & \cellcolor[HTML]{f11f25}\hfil\ding{53} \\
\hline
\end{tabular}
\end{center}
\end{table}

All approximations coincide almost perfectly with the single-Hessian TGA initially: the harmonic approximations during the first 5-10 fs [Fig. S5(a)] and the short-time approximations during the first 3-5 fs [Fig. S5(b)].  At longer times, the discrepancies increase.

Neglecting anharmonicity in the considered time scale (of $\sim$50 fs) would be acceptable for butynal and pentynal, but not for propynal, in which it leads to an artificial coherence revival after 20 fs [compare the global HA and single-Hessian TGA in Fig. S5(a)]. In contrast, including mode mixing (Duschinsky rotation) is important in all three molecules, as the DD HO model significantly underestimates decoherence even in butynal and pentynal. Finally, the importance of including mode distortion is demonstrated by comparing the DD HO model with the DHO model, which neglects mode distortion and exhibits large artificial recoherences, especially in propynal. In conclusion, we believe that anharmonicity should be taken into account to avoid wrong revival and overestimation of the coherence. If only the initial decay is of interest, the use of harmonic models can be justified, but mode distortion and mixing should be included.

All three short-time approximations result in much faster decay of coherence  [Fig. S5(b)]. Among them, only the DR0 permits minor revivals, whereas both STA and STA of DHO impose a complete and irreversible Gaussian decay of coherence. The short-time approximations overestimate decoherence, because the coherence of some, particularly faster, vibrations can increase again after the initial decay, while it is the slower modes that ultimately suppress the coherence (see Fig. 4 of the main text). This also explains why the short-time approximations become less accurate in larger molecules (compare pentynal with butynal and propynal). In contrast to the short-time approximations, the SH-TGA can account for recoherence but does not predict this effect to occur in our studied systems. Yet, recoherence can be important in other systems and, indeed, has been observed experimentally in silane \cite{Matselyukh_Worner:2022}. 

\section*{Assessing nonadiabaticity and nuclear dependence of the electronic wavefunction}
In this section, we justify our neglecting nonadiabatic effects and assuming a weak dependence of the electronic states on nuclear coordinates. Surface-hopping calculations show a noticeable population transfer to the second cationic state in propynal and pentynal, but without significantly affecting charge migration. The second approximation is assessed by evaluating the overlap of the propagated and initial electronic wavefunctions. Our results show that, on the time scale of electronic coherence, the adiabatic states do not change significantly.

\subsection{Energies along semiclassical trajectories}
\begin{figure}[H]
\centering
\includegraphics[]{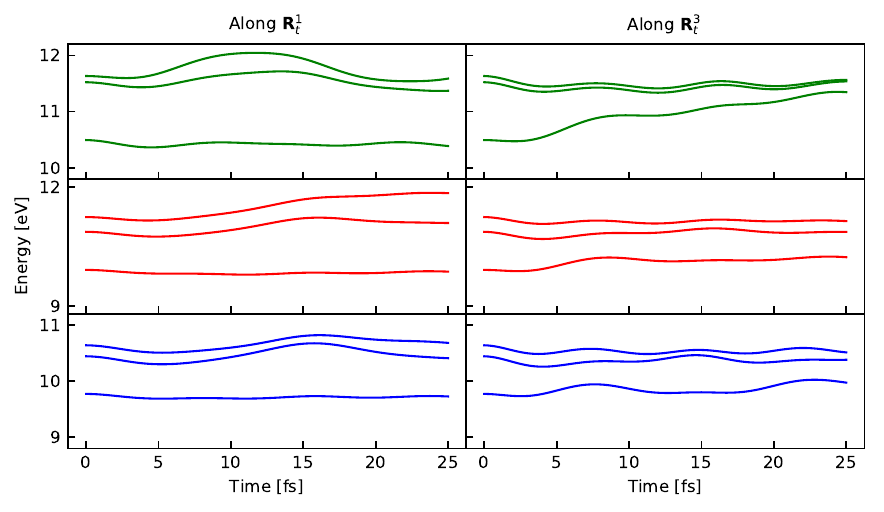}
\caption{Energy of the first three cationic states along the semiclassical trajectory propagated in the first (left) and third (right) cationic states for propynal (top, green), butynal (middle, red) and pentynal (bottom, blue) evaluated at the EOM-CC-IP/DZP level.}
\label{fig:energies_1}
\end{figure}
Energies of the first few cationic states along the two semiclassical trajectories guiding the wavepackets in the involved electronic states are presented in Fig.~\ref{fig:energies_1}. The energy curves never cross, although the third and second electronic states are sometimes very close, especially in propynal, but generally only after the coherence is suppressed by dephasing  (see Fig.~1 of the main text). The first and second cationic states remain well separated.

\subsection{Change of the electronic wavefunction along semiclassical trajectories}
\begin{figure}[H]
\centering
\includegraphics[]{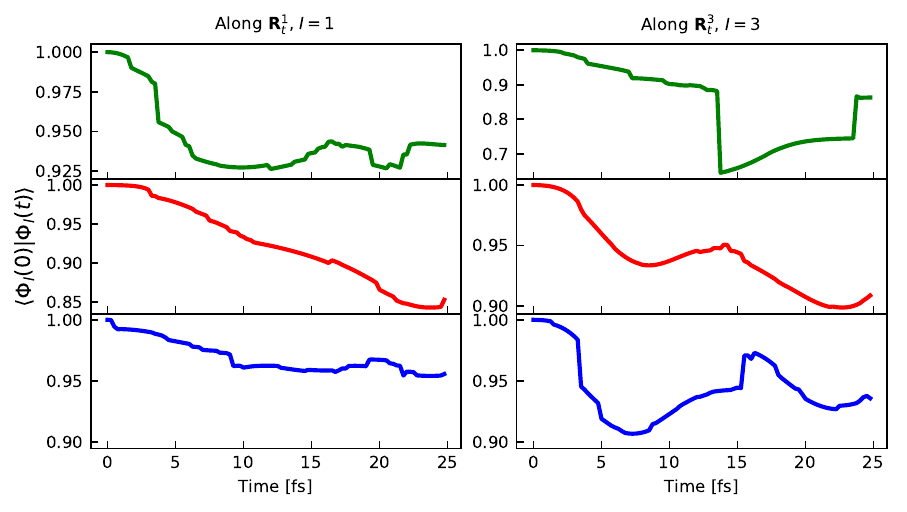}
\caption{Overlap of the electronic wavefunction at time $t$ of the nuclear propagation with the initial wavefunction (at time $0$) for propynal (top, green), butynal (middle, red) and pentynal (bottom, blue). The left panels contain the results for the nuclear propagation on the first cationic surface, and the right ones for the propagation on the third cationic surface.}
\label{fig:MRCI_overlap}
\end{figure}
Our description of charge migration assumes a weak dependence of the electronic wavefunction on nuclear coordinates. To demonstrate the validity of this assumption, we calculated the overlap of the electronic wavefunction along the nuclear trajectory with the initial electronic wavefunction. The evaluation of the overlap of electronic wavefunctions requires an important computational overhead and is not available for all electronic structure methods and in all packages. Fortunately, the MOLPRO 2019 package~\cite{MOLPRO_2019} offers this option for electronic wavefunctions computed with the multireference configuration interaction (MRCI) method. Figure~\ref{fig:MRCI_overlap} shows that the overlap remains above $\sim0.85$ for all three molecules during the first 25 fs, indicating the absence of a sudden change of the adiabatic states. The only exception can be observed for propynal, along the trajectory propagated on the third-state surface. At 14 fs, the electronic wavefunction changes abruptly. However, this sudden change occurs only after the electronic coherence has been suppressed and thus does not change our conclusions.

\subsection{Nonadiabatic dynamics using the fewest-switches surface-hopping algorithm}
\begin{figure}[H]
\centering
\includegraphics[]{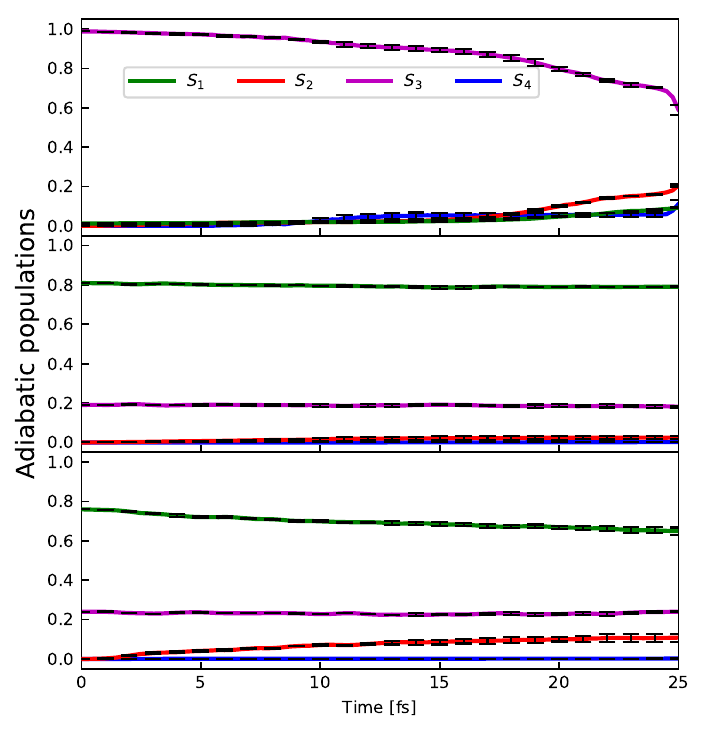}
\caption{Adiabatic populations of the first four cationic states of propynal (top), butynal (middle), and pentynal (bottom) computed with the fewest-switches surface-hopping algorithm.}
\label{fig:SH_population}
\end{figure}

To estimate the extent of nonadiabatic transitions in the three molecules, we performed fewest-switches surface-hopping simulations, in which we propagated $N$ trajectories in Cartesian coordinates, with initial positions and momenta sampled from the harmonic approximation of the neutral ground electronic and nuclear state. The initial populations $(P_{1},P_{3})$ of the first and third electronic states, set according to the hole-mixing amplitudes, were $(1.22\%,98.78\%)$ in propynal, $(81\%,19\%)$ in butynal, and $(76\%,24\%)$ in pentynal. Whereas $N=64$ trajectories were sufficient for both butynal and pentynal, $N=256$ trajectories were needed for propynal because of the small population of the first cationic state. The simulations were conducted twice in order to estimate error bars; the population dynamics plots below display the populations averaged over the two simulations together with error bars.
Figure~\ref{fig:SH_population} shows the electronic populations evaluated from the electronic amplitudes of the trajectories. This choice is motivated by the demonstration by Subotnik \textit{et al.}~\cite{Subotnik_Bellonzi:2016} that the populations evaluated from the electronic amplitudes, rather than the surface occupations, are more accurate for short time dynamics. The ground cationic state was evaluated at the $\omega$B97XD/6-311++G(d,p) level using the density functional theory implemented in the Q-Chem package, and the excited states using the time-dependent analog~\cite{Runge_Gross:1984, Casida:1995}.

\noindent Population transfer is entirely negligible in butynal. In propynal, the population of the third state partially transfers to the second state, with a noticeable acceleration after 20 fs. Similar dynamics occurs in pentynal, with the difference that the increase in the population of the second state starts earlier, is very slow, and is due to a partial depletion of the population of the first (and not the third) cationic state. For all molecules, the fourth electronic state can be neglected, despite a slight increase in its population in propynal after 10 fs.

\noindent Nonadiabatic effects are negligible in propynal and butynal during the coherence time determined by pure dephasing (compare Fig.~\ref{fig:SH_population} with Fig.~1 of the main text). In pentynal, however, the small but early population transfer to the second state could, in principle, affect charge migration. To quantify this, we generalized the expression for the electronic coherence (Eq.~[5] in the main text) to 
\begin{equation}\label{eq:coherence_anl_2}
\chi_{IJ} (t) = a^{*}_{I}(t) a_{J}(t)e^{-d(t)^2/4\hbar} e^{iS(t)/\hbar},
\end{equation}
where we allowed the coefficients $a_{I}(t)$ to be time-dependent and evaluated them approximately from the surface-hopping populations $P_{I}(t)$ as $a_{I}(t)=\sqrt{P_{I}(t)}$.

\begin{figure}[H]
\centering
\includegraphics[]{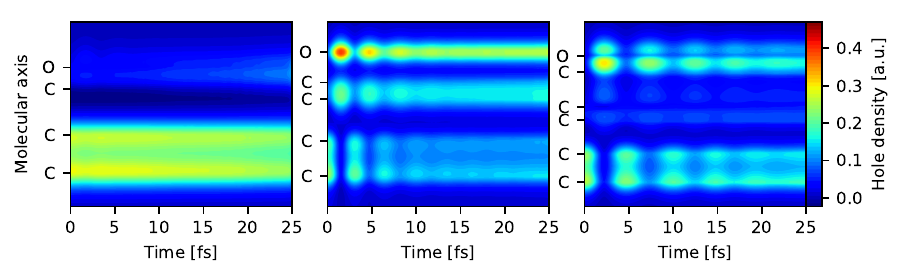}
\caption{Charge migration projected along the molecular axis passing through the carbon triple bond of propynal (left), butynal (middle), and pentynal (right) when nonadiabatic effects are considered.}
\label{fig:charge_mig_w_SH}
\end{figure}

Figure~\ref{fig:charge_mig_w_SH} shows charge migration after ionization when nonadiabatic transitions between the first three (four for propynal) cationic states are included. Comparison with Fig.~1(D) of the main text confirms that nonadiabatic effects are negligible, as the difference between the corresponding densities in the two figures is barely perceptible.
Overall, we can conclude that charge migration in all three molecules is not affected by nonadiabatic effects during the coherence time.

\bibliography{can_increasing_size_and_flexibility_of_molecule_reduce_decoherence}